\documentclass[aps,prl,amsfonts,amssymb,twocolumn,amsmath,preprintnumbers,nofootinbib,floatfix,
showpacs]{revtex4-1}
\usepackage[dvips]{graphics}
\usepackage{graphicx}
\usepackage{bm}
\begin{document}

\title{Recovering of superconductivity in S/F bilayers under spin-dependent nonequilibrium quasiparticle distribution}
\author{I. V. Bobkova}
\affiliation{Institute of Solid State Physics, Chernogolovka,
Moscow reg., 142432 Russia}
\affiliation{Moscow Institute of Physics and Technology, Dolgoprudny, 141700 Russia}
\author{A. M. Bobkov}
\affiliation{Institute of Solid State Physics, Chernogolovka,
Moscow reg., 142432 Russia}

\date{\today}

\begin{abstract}
We study theoretically the influence of spin accumulation on superconductivity in a superconductor/ferromagnet bilayer. 
It is well-known that the superconductivity in S/F bilayers is suppressed by the proximity to a ferromagnet. The spin accumulation by itself is also a depairing factor. But here we show that creation of the spin accumulation on top of effective exchange depairing, caused by the proximity to a ferromagnet, can lead to an opposite result. The superconductivity can be partially recovered by spin-dependent quasiparticle distribution. The systems with realistic parameters are considered and the possible experimental setup is proposed.
\end{abstract}

\maketitle

It is well-known that the Zeeman interaction of electron spins with magnetic or exchange field is destructive to singlet superconductivity. The behavior of a magnetic superconductor with an exchange field $h$ 
was studied long ago \cite{larkin64,fulde64,sarma63,maki68}. It was found that 
homogeneous superconducting state becomes energetically unfavorable above the paramagnetic (Pauli) limit
$h=\Delta/\sqrt 2$. An inhomogeneous state with a spatially 
modulated Cooper pair wave function (LOFF-state) can appear only in a narrow region of exchange 
fields exceeding this value, as it was predicted in \cite{larkin64,fulde64}.

Superconductor/ferromagnet (S/F) hybrid structures also can behave analogous to magnetic superconductors. In particular,
it was shown \cite{bergeret01} that a thin S/F bilayer  is equivalent to a magnetic superconductor in an effective exchange field. Another way to create an exchange field in a thin superconducting
film is to contact it to a ferromagnetic insulator \cite{tedrow86,meservey94,moodera88,hao91,cottet09}, as 
it was observed experimentally \cite{hao91} and justified theoretically \cite{cottet09}.

However, recently it was demonstrated \cite{bobkova11} that the simultaneous applying of the exchange field and creation of spin-dependent quasiparticle distribution in such S/F heterostructures can lead to qualitatively new phenomenon. For a thin superconducting film 
the destructive effect of the exchange field can be fully compensated by the creation of spin-dependent quasiparticle 
distribution in it. This effect takes place even if the exchange field exceeds the paramagnetic limit considerably,
that is under the condition that superconductivity of the equilibrium film is fully suppressed.

In \cite{bobkova11} the effect was illustrated on the basis of a voltage-biased 
half metal/superconductor/half metal (HM/S/HM) heterostructure. 
A thin film (with the thickness less than the superconducting coherence length) is sandwiched between two half-metallic layers with opposite directions of magnetization. 
Half-metallic behavior has been reported in ${\rm CrO}_2$ \cite{soulen98,ji01} 
and in certain manganites \cite{park98}. In-plane effective uniform exchange field $h_{eff}$ 
in the film is supposed to be created by spin-active interfaces with half metals. The spin-dependent 
quasiparticle distribution in the film can be generated by applying a voltage bias between the two half metals.
In this case for spin-up subband the main voltage drop occurs at one of the HM/S interfaces, while for spin-down subband - 
at the other. As a result, the distribution functions for spin-up and spin-down electrons in the superconducting
film are to be close to the equilibrium form with different electrochemical potentials. The superconducting order parameter 
becomes exactly equal to its value for zero exchange field when this difference in electrochemical potentials (spin imbalance) reaches $h_{eff}$. 

For the considered nonequilibrium case the paramagnetic state 
cannot be realized because the distribution function is created and supported by the external conditions
in such a way that the populations of majority and minority subbands in the film remain equal.     

Here we demonstrate that the destructive effect of the exchange field can be compensated by the creation of spin-dependent quasiparticle 
distribution and the superconductivity can be recovered not only for the HM/S/HM heterostructure, proposed in \cite{bobkova11}. The point is that the experimental realization of such a structure is difficult at the moment. First of all, the magnetizations of the half metals should be strictly antiparallel, what is hard to reach experimentally. Second, the effective exchange field, induced in the superconductor, should be of the order of the zero-temperature superconducting gap in order to observe the effect. It also seems to be a problem to get in a controllable way such values of the effective exchange field due to proximity of a half metal. 
 
The effect discussed here is basically the manifestation of the same superconductivity recovering, but it is considered for a system based on a S/F bilayer. The S/F bilayer is a well investigated system as theoretically, so as experimentally. For our purposes it is important that in S/F bilayers there is a mesoscopic analogue of the LOFF-state. This phenomenon  was predicted
theoretically \cite{buzdin82,buzdin92} and observed experimentally \cite{ryazanov01,kontos02,
blum02,guichard03,sidorenko09}.
In this state Cooper pair acquires the total momentum $2Q$ or $-2Q$ inside the ferromagnet as a response
to the energy difference between the two spin directions. Here $Q \propto h/v_F$,
where $h$ is an exchange energy and $v_F$ is
the Fermi velocity. Combination of the two possibilities
results in the spatial oscillations of the condensate wave function $\Psi (x)$ in the ferromagnet along the direction
normal to the SF interface \cite{demler97}. This oscillatory dependence is known to cause $\pi$-Josephson junction formation \cite{buzdin82,ryazanov01} and the non-monotonic (and, in particular, re-entrant) dependence of the critical temperature of S/F bilayers on the F layer thickness \cite{strunk95,jiang95,muhge96,mercaldo96,zdravkov06,zdravkov10}. The effect of superconductivity recovering can be observed in S/F bilayers just in this regime.

In order to create the appropriate spin-dependent distribution function it is enough to contact the S/F bilayer to a strong ferromagnet via a tunnel junction and to pass the electric current through the system. Such a setup is easy to realize experimentally in contrast to the system based on two half metals with strictly opposite magnetizations.

\begin{figure}[!tbh]
  \centerline{\includegraphics[clip=true,width=2.7in]{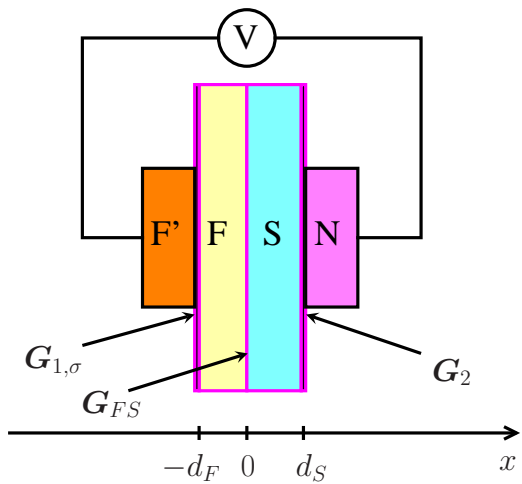}}
           %\centerline{\includegraphics[clip=true,width=2.5in]{fig1b.eps}}
  %\begin{minipage}[b]{0.5\linewidth}
   %  \centerline{\includegraphics[clip=true,width=1.5in]{fig1a.eps}}
    % \end{minipage}\hfill
    %\begin{minipage}[b]{0.5\linewidth}
   %\centerline{\includegraphics[clip=true,width=1.5in]{fig1b.eps}}
  %\end{minipage}
%\begin{minipage}[b]{0.5\linewidth}
 %    \centerline{\includegraphics[clip=true,width=1.5in]{fig1c.eps}}
  %   \end{minipage}\hfill
   % \begin{minipage}[b]{0.5\linewidth}
   %\centerline{\includegraphics[clip=true,width=1.5in]{fig1d.eps}}
  %\end{minipage}
   \caption{Sketch of the system under consideration}   
\label{system}
\end{figure}

Now we turn to the detailed description of the proposed system and to the microscopic calculation. The sketch of the system under consideration in represented in Fig.~\ref{system}. The S/F bilayer is a main part of the setup. It is composed of a singlet s-wave superconductor S and a weak ferromagnetic alloy F with the thicknesses $d_S$ and $d_F$, respectively. The $x$-axis is normal to the bilayer plane and the F/S interface is at $x=0$. The bilayer is sandwiched between the normal metal N and a strong ferromagnet F' (Fe,Ni,Co) via tunnel junctions. The system is biased by the voltage $V$ in order to create the spin-dependent nonequilibrium distribution in the bilayer. 

In our calculations we assume that (i) the system is in the dirty limit, so the quasiclassical Green's function obeys Usadel equations \cite{usadel}; (ii) the thickness of the S layer $d_S \lesssim \xi_S$. Here $\xi_S=\sqrt{D_S/\Delta_0}$ is the superconducting coherence length, $D_S$ is the diffusion constant in the superconductor and $\Delta_0$ is the bulk value of the superconducting order parameter at zero temperature. This condition allows us to neglect the variations of the superconducting order parameter and the Green's functions across the S layer; (iii) we work in the vicinity of the critical temperature, so the Usadel equations can be linearized with respect to the anomalous Green's function.

The retarded anomalous Green's function $\hat f^R(\varepsilon,x)$ is a $2\times 2$ matrix in spin space. We assume that the exchange field in the F layer is homogeneous $\bm h=(0,0,h)$. In this case there are only singlet and triplet with zero spin projection on the quantization axis pairs in the system. In the language of Pauli matrices it means that $\hat f^R(\varepsilon,x)=[f_\uparrow^R (1+\sigma_3)/2+f_\downarrow^R (1-\sigma_3)/2]i\sigma_2$, where $\sigma_{2,3}$ are the corresponding Pauli matrices in spin space. While we only consider the singlet pairing channel, the superconducting order parameter $\hat \Delta=\Delta i \sigma_2$.

The linearized Usadel equation for the retarded anomalous Green's function $f^R_\sigma$, where $\sigma=\uparrow,\downarrow$, takes the form:
\begin{equation}
D \partial_x^2 f^R_\sigma + 2 i (\varepsilon+\sigma h(x))  f^R_\sigma -2 i \Delta(x) = 0
\enspace .
\label{usadel}
\end{equation}
Here $\sigma = \pm 1$ for $f_{\uparrow(\downarrow)}$. $D$ stands for the diffusion constant, which is equal to $D_{S(F)}$ in the superconductor (ferromagnet). $h(x) = h$ in the ferromagnet and $h(x)=0$ in the superconductor. Analogously, $\Delta(x)=0$ in the ferromagnet and $\Delta(x)=\Delta$ in the superconductor. 

Eq.~(\ref{usadel}) should be supplied by the Kupriyanov-Lukichev boundary conditions \cite{kupriyanov88} at the S/F interface ($x=0$):
\begin{eqnarray}
\sigma_S \partial_x f^R_{\sigma,S} = \sigma_F \partial_x f^R_{\sigma,F} = G_{FS}\left.(f^R_{\sigma,S}-f^R_{\sigma,F})\right|_{x=0}
\label{interface_cond}
\enspace ,
\end{eqnarray}
where $\sigma_{S(F)}$ stands for a conductivity of the S(F) layer and $G_{FS}$ is the conductance of the S/F interface. The boundary conditions at the ends of the bilayer are $\left. \partial_x f^R_{\sigma,S} \right|_{x=d_S} = \left. \partial_x f^R_{\sigma,F} \right|_{x=-d_F}=0 $. Here we neglect small conductances $G_{1,2}$ of the F'/F and S/N interfaces because they enter the resulting anomalous Green's function only as very small additional deparing factors.  

Solving Eq.~(\ref{usadel}) under the assumption, that the anomalous Green's function weakly varies across the S layer, we obtain the anomalous Green's functions in the bilayer. In the S layer it take the form:
\begin{equation}
f_{\sigma,S}^R=\frac{\Delta}{E}
\label{fs}
\enspace ,
\end{equation}
\begin{equation} 
E=\varepsilon+ \frac{i G_{FS}D_S \lambda_\sigma \tanh[\lambda_\sigma d_F]}{2\sigma_S d_S (\lambda_\sigma \tanh[\lambda_\sigma  d_F]+G_{FS}/\sigma_F)} 
\label{E}
\enspace ,
\end{equation}
where $\lambda_\sigma^2=-2i(\varepsilon+\sigma h)/D_F $.

\begin{figure}[!tbh]
  \centerline{\includegraphics[clip=true,width=2.7in]{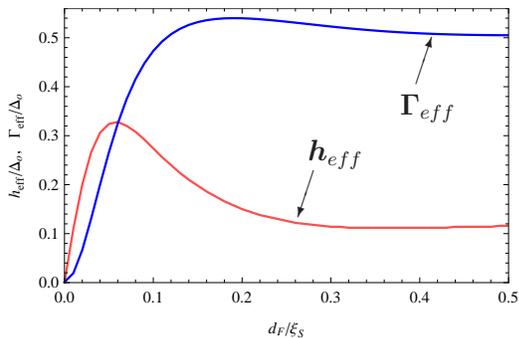}}
           %\centerline{\includegraphics[clip=true,width=2.5in]{fig1b.eps}}
  %\begin{minipage}[b]{0.5\linewidth}
   %  \centerline{\includegraphics[clip=true,width=1.5in]{fig1a.eps}}
    % \end{minipage}\hfill
    %\begin{minipage}[b]{0.5\linewidth}
   %\centerline{\includegraphics[clip=true,width=1.5in]{fig1b.eps}}
  %\end{minipage}
%\begin{minipage}[b]{0.5\linewidth}
 %    \centerline{\includegraphics[clip=true,width=1.5in]{fig1c.eps}}
  %   \end{minipage}\hfill
   % \begin{minipage}[b]{0.5\linewidth}
   %\centerline{\includegraphics[clip=true,width=1.5in]{fig1d.eps}}
  %\end{minipage}
   \caption{Effective exchange field $h_{eff}$ and depairing factor $\Gamma_{eff}$ as functions of $d_F$. The particular parameters are the following: $d_S=0.36\xi_S$, $h=5 \Delta_0$, $\frac{G_{FS}\xi_S}{\sigma_S}=0.5$, $D_S/D_F=5$ and $\sigma_S/\sigma_F=6$.}   
\label{depairing}
\end{figure}

Due to the fact that the bilayer is thin as compared to the superconducting coherence length, the anomalous Green's function in it  takes the form of Eq.~(\ref{fs}), characteristic for a homogeneous superconductor. The denominator $E$ of Eq.~(\ref{fs}) can be approximately represented as $\varepsilon + \sigma h_{eff} + i\Gamma_{eff}$, where the effective exchange field $h_{eff}$ and the depairing factor $\Gamma_{eff}$ are caused by the proximity of the S film to the ferromagnet. The dependence of $h_{eff}$ and $\Gamma_{eff}$ is represented in Fig.~\ref{depairing}.

\begin{figure}[!tbh]
  %\centerline{\includegraphics[clip=true,width=2.7in]{Fig3.eps}}
           %\centerline{\includegraphics[clip=true,width=2.5in]{fig1b.eps}}
  \begin{minipage}[b]{\linewidth}
     \centerline{\includegraphics[clip=true,width=2.7in]{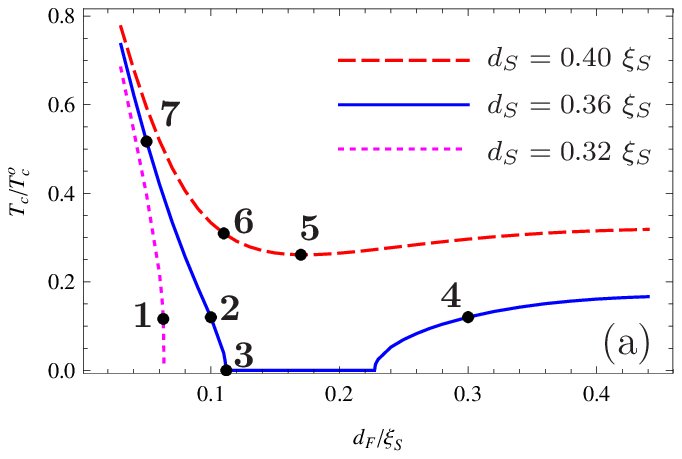}}
     \end{minipage}\hfill
    \begin{minipage}[b]{\linewidth}
   \centerline{\includegraphics[clip=true,width=2.7in]{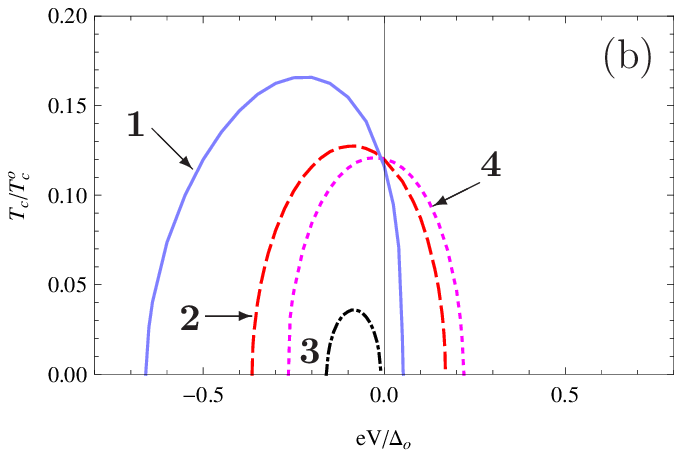}}
  \end{minipage}
\begin{minipage}[b]{\linewidth}
     \centerline{\includegraphics[clip=true,width=2.7in]{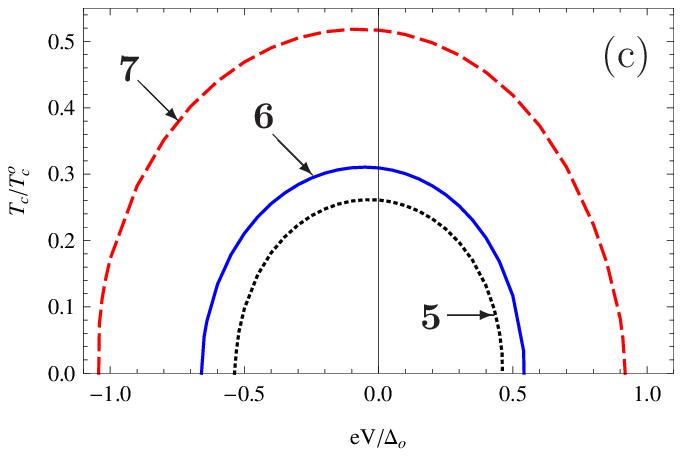}}
     \end{minipage}\hfill
   % \begin{minipage}[b]{0.5\linewidth}
   %\centerline{\includegraphics[clip=true,width=1.5in]{fig1d.eps}}
  %\end{minipage}
   \caption{(a) Critical temperature of the S/F bilayer as a function of $d_F$ for different values of $d_S$. (b)-(c) Critical temperature of the bilayer as a function of $V$. Different curves in panels (b) and (c) are calculated for the bilayers with particular $d_F$ and $d_S$, corresponding to the same numbers in panel (a). The other parameters are the same as in Fig.~\ref{depairing}. The temperature is normalized to the critical temperature of the superconducting film $T_{c0}$ in the absence of a ferromagnet and $V$ is normalized to $\Delta_0$. All the lengths are in units of $\xi_S$.}   
\label{Tc}
\end{figure}

The critical temperature of the bilayer should be calculated from the self-consistency equation. For the equilibrium case it takes the form
\begin{equation} 
\Delta=\Lambda \int \limits_{-\omega_D}^{\omega_D} \frac{d \varepsilon}{4} \sum \limits_\sigma {\rm Re}\left[ f_{\sigma,S}^R \right]\tanh \frac{\varepsilon}{2T_c}
\label{self_eq}
\enspace ,
\end{equation}
where $\Lambda$ is a dimensionless pairing constant. The calculated dependence of the critical temperature on $d_F$ is shown in Fig.~\ref{Tc}(a). Different curves correspond to different values of $d_S$. It is seen that for very thin S films with $d_S<d_{S,cr}$ the critical temperature is simply suppressed upon increase of $d_F$. However, there is a range of $d_S>d_{S,cr}$, where $T_c$ manifests a nonmonotonous and, even, a re-entrant dependence on $d_F$. These equilibrium results are well-known and were reported in the literature as theoretically \cite{khusainov00,tagirov98,vodopyanov03}, so as experimentally \cite{strunk95,jiang95,muhge96,mercaldo96,zdravkov06,zdravkov10}.

\begin{figure}[!tbh]
  \centerline{\includegraphics[clip=true,width=2.7in]{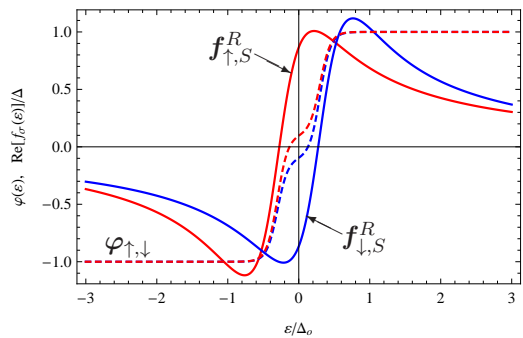}}
           %\centerline{\includegraphics[clip=true,width=2.5in]{fig1b.eps}}
  %\begin{minipage}[b]{0.5\linewidth}
   %  \centerline{\includegraphics[clip=true,width=1.5in]{fig1a.eps}}
    % \end{minipage}\hfill
    %\begin{minipage}[b]{0.5\linewidth}
   %\centerline{\includegraphics[clip=true,width=1.5in]{fig1b.eps}}
  %\end{minipage}
%\begin{minipage}[b]{0.5\linewidth}
 %    \centerline{\includegraphics[clip=true,width=1.5in]{fig1c.eps}}
  %   \end{minipage}\hfill
   % \begin{minipage}[b]{0.5\linewidth}
   %\centerline{\includegraphics[clip=true,width=1.5in]{fig1d.eps}}
  %\end{minipage}
  \caption{Characteristic behavior of the distribution function (dashed lines) and the anomalous Green's function (solid lines) in dependence on the quasiparticle energy. Different spin subbands are denoted by blue and red lines.}   
\label{distrib}
\end{figure}

Now let us insert a S/F bilayer, described above, into the setup, depicted in Fig.~\ref{system} and apply a voltage bias $V$ to it. The anomalous Green's function in the S film is still determined by Eq.~(\ref{fs}), but the distribution function in the bilayer is now strongly non-equilibrium and should be determined from the Keldysh part of the Usadel equation and Kupriyanov-Lukichev boundary conditions. The distribution function is a $4 \times 4$ matrix in the direct product of spin and particle-hole spaces. It is always diagonal in particle-hole space: $\check \varphi=\hat \varphi (1+\tau_3)/2+\hat {\tilde \varphi} (1-\tau_3)/2$. The electron $\hat \varphi$ and hole $\hat {\tilde \varphi}$ components of the distribution function are $2 \times 2$ matrices in spin space. Here we suppose that the exchange field of the weak ferromagnetic alloy F is aligned with the magnetization of the strong ferromagnet F', so in our problem we have the only magnetization direction. We choose the quantization axis along this direction. Then the distribution function is diagonal in spin space $\hat \varphi = \left(\begin{array}{cc} \varphi_\uparrow & 0 \\ 0 & \varphi_\downarrow   \end{array}\right)$ and the general symmetry relation between the electron and hole parts of the distribution function takes the form $\tilde \varphi_{\bar \sigma}(\varepsilon)=-\varphi_\sigma (-\varepsilon)$, where $\bar \sigma$ denotes the spin direction opposite to $\sigma$. For our linearized problem the kinetic equation takes the same form as in the normal metal:
\begin{equation}
D_{S,F} \partial_x^2 \varphi_\sigma + S[\hat \varphi]=0
\label{kin}
\enspace ,
\end{equation}
where $S[\hat \varphi]$ is a collision term due to energy and spin relaxation processes in the bilayer. Further we will assume that $\tau_{sf,\varepsilon}^{-1} \ll \frac{D_F G_1}{(d_F+d_S)\sigma_F},\frac{D_S G_2}{(d_F+d_S)\sigma_S}$ and $\frac{G_1 d_F}{\sigma_F},\frac{ G_2 d_S}{\sigma_S} \ll 1$. The first inequality means that the relaxation processes ($\tau_{sf,\varepsilon}$ accounts for the spin and energy relaxation, respectively) are weak and the distribution function in the bilayer is mainly determined by the interchange with the reservoirs. The second inequality means that the electric current through the bilayer is very small, so the main voltage drop occurs at the tunnel F'/F and S/N interfaces and the distribution function is approximately constant over the thickness of the bilayer. Under these assumptions the distribution function has a double step structure
\begin{equation}
\varphi_\sigma=\varphi_1 \frac{G_{1,\sigma}}{G_{1,\sigma}+G_2}+\varphi_2 \frac{G_2}{G_{1,\sigma}+G_2}
\label{varphi}
\enspace .
\end{equation}
Here $\varphi_{1,2}=\tanh[(\varepsilon-eV_{1,2})/2T]$ are the distribution functions of left (F') and right (N) reservoirs, which are assumed to have the equilibrium form shifted by $V_{1,2}$. $G_2$ is the conductance of the S/N interface and $G_{1,\sigma}$ is the conductance of the F'/F interface for a spin $\sigma$. Typical experimental values of the polarization $P=(G_\uparrow -G_\downarrow )/(G_\uparrow +G_\downarrow)$ of a tunnel interface between a strong ferromagnet and nonmagnetic material are of the order of $0.1-0.2$ \cite{hubler12,wolf13,wolf14}, so we assume in our calculation $P=0.2$. The conductances are taken to be $G_2^2=G_{1,\uparrow}G_{1,\downarrow}$. In this symmetric case $\tilde \varphi_\sigma=\varphi_\sigma$ and the self-consistency equation takes the form
\begin{equation} 
\Delta=\Lambda \int \limits_{-\omega_D}^{\omega_D} \frac{d \varepsilon}{4} \sum \limits_\sigma {\rm Re}\left[ f_{\sigma,S}^R \right]\varphi_\sigma
\label{self_non}
\enspace .
\end{equation}
The typical behavior of the distribution function described by Eq.~(\ref{varphi}) is shown in Fig.~\ref{distrib} as a function of the quasiparticle energy. The anomalous Green's function for the both spin subbands is also represented in this figure.

In Figs.~\ref{Tc}(b) and \ref{Tc}(c) the critical temperature of the nonequilibrium S/F bilayer is demonstrated as a function of the voltage applied between the external electrodes. These are the central results of the paper. It is seen that $T_c$ is an asymmetric function of $V$. Possibly, in real experiment it is more convenient to fix the temperature and to study if the superconductivity is completely suppressed at equal positive and negative voltages or not.

We have checked that if the distribution function is spin-independent ($P=0$), then $T_c$ is suppressed symmetrically as a function of $V$. It is also known that if there is no effective exchange field in the superconductor ($h_{eff}=0$), the spin-dependent distribution simply suppresses superconductivity \cite{takahashi99,vasko97,dong97}. So, the results, presented here, is a manifestation of superconductivity recovering under spin-dependent quasiparticle distribution, predicted in \cite{bobkova11}. 

It is worth to note that for the S/F system we cannot get the "full recovering" of superconductivity up to $T_{c0}$, in contrast to \cite{bobkova11}. The reason is seen from Fig.~\ref{distrib}. Here the distribution function is not a "pure spin imbalance", as in \cite{bobkova11}, but it contains a small spin imbalance on the background of strong spin-independent nonequilibrium, which just suppresses superconductivity. So, in our system two opposite effects act simultaneously: the spin-dependent part of the distribution recovers superconductivity, and the spin-independent part tends to suppress it. Due to this reason the value of the voltage $V$, which provides the maximal $T_c$ for a given sample, is considerably smaller than the effective exchange $h_{eff}$ in this sample. The second thing is that in the bilayer the equilibrium superconductivity is typically suppressed by two factors: $h_{eff}$ and $\Gamma_{eff}$ (see Fig.~\ref{depairing}). The spin-dependent distribution is able to compensate the suppression caused by the effective exchange, but cannot compensate the part of suppression caused by $\Gamma_{eff}$.  

We can conclude that in order to observe the essential effect in a S/F bilayer, (i) it should be close to the regime of re-entrant superconductivity (but not necessary in this regime). In this case we can guarantee that the superconductivity is suppressed, at least partially, by the effective exchange, caused by the proximity to a ferromagnet. (ii) Thinner ferromagnets are more preferable than the thick ones [compare curves marked by 2 and 4 in Fig.~\ref{Tc}(b)]. The reason is that $\Gamma_{eff}$ grows considerably for thicker ferromagnets, as shown in Fig.~\ref{depairing} and becomes the dominating suppressing factor. (iii) The effect is most pronounced if the superconductivity is already strongly suppressed by the proxomity to a ferromagnet, but $\Gamma_{eff}$ is not very large yet [point 1 in Fig.~\ref{Tc}(a) and the corresponding curve in panel (b)]. In this case the main suppressing factor is $h_{eff}$ and it can be partially compensated by the spin-dependent part of the quasiparticle distribution. In principle, the superconductivity can be recovered even in the region of full suppression [point 3 in Fig.~\ref{Tc}(a)], but the spin-dependent part of our distribution is not enough to recover the superconductivity deeply in this region. 

The regimes of weak superconductivity suppression [point 7 in Fig.~\ref{Tc}(a)] are also bad for observation of the effect because $h_{eff}$ can be compensated only partially and the resulting increase of $T_c$ is too small.

In summary, on the basis of the S/F bilayer we study the nonequilibrium recovering of superconductivity, suppressed by the exchange field, and propose a realistic setup for experimental investigation of this effect.

 %\begin{figure}[!tbh]
  %\centerline{\includegraphics[clip=true,width=3.5in]{fig4.eps}}
           %\centerline{\includegraphics[clip=true,width=2.5in]{fig1b.eps}}
  %\begin{minipage}[b]{0.5\linewidth}
   %  \centerline{\includegraphics[clip=true,width=1.5in]{fig1a.eps}}
    % \end{minipage}\hfill
    %\begin{minipage}[b]{0.5\linewidth}
   %\centerline{\includegraphics[clip=true,width=1.5in]{fig1b.eps}}
  %\end{minipage}
%\begin{minipage}[b]{0.5\linewidth}
 %    \centerline{\includegraphics[clip=true,width=1.5in]{fig1c.eps}}
  %   \end{minipage}\hfill
   % \begin{minipage}[b]{0.5\linewidth}
   %\centerline{\includegraphics[clip=true,width=1.5in]{fig1d.eps}}
  %\end{minipage}
  % \caption{}   
%\label{fig}
%\end{figure}

{\it Acknowledgments.} The work was supported by RFBR Grant No. 12-02-00723.

%\begin{figure}[!tbh]
 % \centerline{\includegraphics[clip=true,width=3.5in]{fig4.eps}}
            %\centerline{\includegraphics[clip=true,width=2.5in]{fig1b.eps}}
   %\begin{minipage}[b]{\linewidth}
    % \centerline{\includegraphics[clip=true,width=2.7in]{fig2a.eps}}
     %\end{minipage}\hfill
    %\begin{minipage}[b]{\linewidth}
   %\centerline{\includegraphics[clip=true,width=2.64in]{fig2b.eps}}
  %\end{minipage}
  % \caption{}   
%\label{distrib_dependence}
%\end{figure}

%\bibliography{/users/tkm/howell/latexx/bibtexx/refs}

\begin{thebibliography}{99}
%
\bibitem{larkin64}
A.I. Larkin and Yu.N. Ovchinnikov, Sov. Phys. JETP {\bf 20}, 762 (1965) [Zh. Eksp. Teor. Fiz. {\bf 47}, 1136 (1964)].
%
\bibitem{fulde64}
P. Fulde and R.A. Ferrel, Phys.Rev. {\bf 135}, A550 (1964).
%
\bibitem{sarma63}
G. Sarma, J. Phys. Chem. Solids {\bf 24}, 1029 (1963).
%
\bibitem{maki68}
K. Maki, Progr. Theoret. Phys. {\bf 39}, 897 (1968).
%
\bibitem{bergeret01}
F.S. Bergeret, A.F. Volkov, and K.B. Efetov, Phys. Rev. Lett. {\bf 86}, 3140 (2001).
%
\bibitem{tedrow86}
P.M. Tedrow, J.E. Tkaczyk, and A. Kumar, Phys. Rev. Lett. {\bf 56}, 1746 (1986).
%
\bibitem{meservey94}
R. Meservey and P.M. Tedrow, Phys. Rep. {\bf 238}, 173 (1994).
%
\bibitem{moodera88}
J.S. Moodera, X. Hao, G.A. Gibson, and R. Meservey, Phys. Rev. Lett. {\bf 61}, 637 (1988).
%
\bibitem{hao91}
X. Hao, J.S. Moodera, and R. Meservey, Phys. Rev. Lett. {\bf 67}, 1342 (1991).
%
\bibitem{cottet09}
A. Cottet, D. Huertas-Hernando, W. Belzig, and Yu.V. Nazarov, Phys. Rev. B {\bf 80}, 184511 (2009).
%
\bibitem{bobkova11}
I.V. Bobkova and A.M. Bobkov, Phys. Rev. B {\bf 84}, 140508(R) (2011).
%
\bibitem{soulen98}
R. J. Soulen, Jr., J. M. Byers, M. S. Osofsky, B. Nadgorny, T. Ambrose, S. F. Cheng, P. R. Broussard, C. T. Tanaka, J. Nowak, J. S. Moodera, A. Barry, and J. M. D. Coey, Science {\bf 282}, 85 (1998).
%
\bibitem{ji01}
Y. Ji, G. J. Strijkers, F. Y. Yang, C. L. Chien, J. M. Byers, A. Anguelouch, Gang Xiao, and A. Gupta, Phys. Rev. Lett. {\bf 86},  5585 (2001).
%
\bibitem{park98}
J.-H. Park,  E. Vescovo,  H.-J. Kim,  C. Kwon,  R. Ramesh, and  T. Venkatesan, Nature (London) {\bf 392}, 794 (1998).
%
\bibitem{buzdin82}
A.~I.~Buzdin, L.~N.~Bulaevsky, and S.~V.~Panyukov, JETP Lett. {\bf
35}, 178 (1982) [Pis'ma~Zh.~Eksp.~Teor.~Fiz. {\bf 35}, 147
(1982)].
%
\bibitem{buzdin92}
A.~I.~Buzdin, B.~Bujicic, and M.~Yu.~Kupriyanov, Sov.~Phys.~JETP
{\bf 74}, 124 (1992) [Zh.~Eksp.~Teor.~Fiz. {\bf 101}, 231 (1992)].
%
\bibitem{ryazanov01}
V.~V.~Ryazanov, V.~A.~Oboznov, A.~Yu.~Rusanov, A.~V.~Veretennikov,
A.~A.~Golubov, and J.~Aarts, Phys. Rev. Lett. {\bf 86}, 2427
(2001).
%
\bibitem{kontos02}
T.~Kontos, M.~Aprili, J.~Lesueur, F.~Genet, B.~Stephanidis,
R.~Boursier, Phys. Rev. Lett. {\bf 89}, 137007
(2002).
%
\bibitem{blum02}
Y. Blum, A. Tsukernik, M. Karpovski, and A. Palevski, Phys. Rev. Lett. {\bf 89}, 187004
(2002).
%
\bibitem{guichard03}
W. Guichard, M.Aprili, O. Bourgeois, T. Kontos, J. Lesueur, and P. Gandit, Phys. Rev. Lett. {\bf 90}, 167001
(2003).
%
\bibitem{sidorenko09}
A.S. Sidorenko, V.I. Zdravkov, J. Kehrle, R. Morari, G. Obermeier, S. Gsell, M. Schreck,
C. M${\rm \ddot u}$ller, M.Yu. Kupriyanov, V.V. Ryazanov, S. Horn, L.R. Tagirov, R. Tidecks, JETP Lett. {\bf
90}, 139 (2009) [Pis'ma~Zh.~Eksp.~Teor.~Fiz. {\bf 90}, 149
(2009)].
%
\bibitem{demler97}
E.~A.~Demler, G.~B.~Arnold, M.~R.~Beasley, Phys. Rev. B {\bf 55},
15174 (1997).
%
\bibitem{strunk95}
C. Strunk, C. Surgers, U. Paschen, and H. v. Lohneysen, Phys. Rev. B {\bf 49}, 4053 (1994).
%
\bibitem{jiang95}
J.S. Jiang, D. Davidovic, Daniel H. Reich, and C.L. Chien, Phys. Rev. Lett. {\bf 74}, 314 (1995).
%
\bibitem{muhge96}
Th. Muhge, N. N. Garifyanov,Yu. V. Goryunov, G. G. Khaliullin, L. R. Tagirov,
K. Westerholt, I. A. Garifullin, and H. Zabel, Phys. Rev. Lett. {\bf 77}, 1857 (1996).
%
\bibitem{mercaldo96}
L. V. Mercaldo, C. Attanasio, C. Coccorese, L. Maritato, S. L. Prischepa, and
M. Salvato, Phys. Rev. B 53, 14040 (1996).
% 
\bibitem{zdravkov06}
V. Zdravkov, A. Sidorenko, G. Obermeier, S. Gsell, M. Schreck, C. Muller, S. Horn, R. Tidecks, and L. R. Tagirov, Phys. Rev. Lett. {\bf 97}, 057004 (2006).
%
\bibitem{zdravkov10}
V.I. Zdravkov, J. Kehrle, G. Obermeier, S. Gsell, M. Schreck, C. M$\rm {\ddot u}$ller, H. A. Krug von Nidda, J. Lindner, J. Moosburger-Will, E. Nold, R. Morari, V.V. Ryazanov, A.S. Sidorenko, S. Horn, R. Tidecks, and L.R. Tagirov Phys. Rev. B {\bf 82}, 054517 (2010).
%
\bibitem{usadel}
K.D. Usadel, Phys.Rev.Lett. {\bf 25}, 507 (1970).
%
\bibitem{kupriyanov88}
M.Yu. Kuprianov and V.F. Lukichev, Sov. Phys. JETP {\bf 67}, 1163
(1988).
%
\bibitem{khusainov00}
M. G. Khusainov and Yu. N. Proshin, Phys. Rev. B {\bf 56}, R14283 (1997); Erratum: Phys. Rev. B {\bf 62}, 6832 (2000).
%
\bibitem{tagirov98}
L. R. Tagirov, Physica C {\bf 307}, 145 (1998).
%
\bibitem{vodopyanov03}
B. P. Vodopyanov and L. R. Tagirov, Pis'ma v ZhETF  {\bf 78}, 1043 (2003) [JETP Letters {\bf 78}, 555 (2003)].
%
\bibitem{hubler12}
F. Hubler, M.J. Wolf, D. Beckmann, and H.v. Lohneysen, Phys. Rev. Lett. {\bf 109}, 207001
(2012).
%
\bibitem{wolf13}
M.J. Wolf, F. Hubler, S. Kolenda, H.v. Lohneysen, and D. Beckmann, Phys. Rev. B {\bf 87}, 024517
(2013).
%
\bibitem{wolf14}
M.J. Wolf, C. Surgers, G. Fisher, and D. Beckmann, Phys. Rev. B {\bf 90}, 144509 (2014).
%
\bibitem{takahashi99}
S. Takahashi, H. Imamura, and S. Maekawa, Phys. Rev. Lett. {\bf 82}, 3911 (1999).
%
\bibitem{vasko97}
V. A. Vas'ko, V. A. Larkin, P. A. Kraus, K. R. Nikolaev, D. E. Grupp, C. A. Nordman, and A. M. Goldman, Phys. Rev. Lett. {\bf 78}, 1134 (1997).
%
\bibitem{dong97}
Z. W. Dong, R. Ramesh, T. Venkatesan, Mark Johnson, Z. Y. Chen, S.P. Pai, V. Talyansky, R.P. Sharma, R. Shreekala, C.J. Lobb and R.L. Greene, Appl. Phys. Lett. {\bf 71}, 1718 (1997).
%
\end{thebibliography}

\end{document}